\documentclass[12pt,preprint]{aastex}

\usepackage{colordvi}

\begin{document}
\title{Sudden Photospheric Motion and Sunspot Rotation Associated with the X2.2 Flare on 2011 February 15}
\author{Shuo Wang, Chang Liu, Na Deng, and Haimin Wang}
\affil{Space Weather Research Laboratory, New Jersey Institute of Technology,\\University Heights, Newark, NJ 07102-1982, USA}
\email{shuo.wang@njit.edu}

\begin{abstract}

The Helioseismic and Magnetic Imager provides 45 s cadence intensity images and 720 s cadence vector magnetograms. These unprecedented high-cadence and high-resolution data give us a unique opportunity to study the change of photospheric flows and sunspot rotations associated with flares. By using the differential affine velocity estimator method and the Fourier local correlation tracking method separately, we calculate velocity and vorticity of photospheric flows in the flaring NOAA AR 11158, and investigate their temporal evolution around the X2.2 flare on 2011 February 15. It is found that the shear flow around the flaring magnetic polarity inversion line exhibits a sudden decrease, and both of the two main sunspots undergo a sudden change in rotational motion during the impulsive phase of the flare. These results are discussed in the context of the Lorentz-force change that was proposed by \cite{2008ASPC..383..221H} and \cite{2012SoPh..277...59F}. This mechanism can explain the connections between the rapid and irreversible photospheric vector magnetic field change and the observed short-term motions associated with the flare. In particular, the torque provided by the horizontal Lorentz force change agrees with what is required for the measured angular acceleration.

\end{abstract}


\section{Introduction}

Besides the magnetic complexity and its evolution in flare productive active regions, the flow field is another important factor contributing to the energy storage and release of solar eruptions. \cite{1973SoPh...32..365M,1982PASJ...34..299M} studied photospheric vortex motions, and found that they are correlated with the magnetic field evolution. Recently, rotating sunspots and other magnetic structures have been studied using space-borne observations with high spatial and temporal resolution. \cite{2003SoPh..216...79B} analyzed rotations of seven spots by examining feature movement on uncurled penumbral time-slices. The authors suggested that the rotation may be due to magnetic flux tube emergence. \cite{1993SoPh..147..287A} observed rapid translational motion of a sunspot associated with the 1991 November 15 X1.1 flare, and suggested that the horizontal Lorentz force change can be sufficient to drive the sunspot motion. Closely related horizontal flow motion along the magnetic polarity inversion line (PIL) was found in the photosphere and chromosphere by \cite{1976SoPh...47..233H}. \cite{2004ApJ...617L.151Y} analyzed photospheric shear flows in the NOAA AR 10486 and related them to the flare occurrence. \cite{2009ApJ...690.1820T} found that the shear flow along the PIL dropped by 50\% after a major flare. \cite{2012AN....333..125B} compared the flow field before and after the 2011 February 15 X2.2 flare, and revealed that the shear flow around the flaring PIL changed after the flare.

It is well known that the observed surface flow field is coupled to the evolution of the photospheric magnetic field. Rapid and permanent flare-related changes of magnetic fields on the photosphere in terms of magnetic shear were discovered by \cite{1992SoPh..140...85W} and \cite{1994ApJ...424..436W}. The change in the line-of-sight field component was also recognized \citep[e.g.,][]{2002ApJ...576..497W,2002ApJ...572.1072S, 2004ApJ...605..546Y,2005ApJ...635..647S,2006ApJ...649..490W, 2010ApJ...724.1218P,2013SoPh..283..429B}. \cite{2010ApJ...716L.195W} suggested that the vector magnetic field changes are mainly in the form of the horizontal field enhancement at flaring PILs. In the mean time, the permanent intensity change in the penumbral and umbral regions related to the magnetic field change was also revealed \citep{2004ApJ...605..931W,2005ApJ...623.1195D,2005ApJ...622..722L, 2007ChJAA...7..733C,2008ApJ...676L..81J,2012ApJ...748...76W, 2013ApJ...774L..24W}. Most recently, the photospheric magnetic field change after flares was reconfirmed with space-based vector magnetogram observations \citep[e.g.,][]{2012ApJ...745L...4L,2012ApJ...745L..17W,2012ApJ...757L...5W,2012ApJ...748...77S,liu13}. \cite{2012ApJ...745L..17W} found that there was a rapid enhancement of the horizontal magnetic field in a compact region along the PIL of the 2011 February 15 X2.2 flare. \cite{2013SoPh..287..415P} studied the azimuthal change of horizontal magnetic field in the main spots of this active region during the X2.2 flare, and further pointed out the co-spatial torsional Lorentz force change as well as the sheared Lorentz force along the PIL.

From the theoretical point of view, \cite{2008ASPC..383..221H} used the vertical component of Lorentz force change to assess the back reaction of the coronal restructuring expected from a more horizontal photospheric magnetic field after flares. \cite{2012SoPh..277...59F} formulated the changes of both the vertical and horizontal Lorentz forces implied by the observed changes of vector magnetic fields associated with flares, and further discussed the back reaction scenario using the principle of momentum and energy conservation. The idea of  back reaction has already been reflected in some flare models. For example, the tether-cutting reconnection model for sigmoids \citep{2001ApJ...552..833M} suggests two new flux loops associated with an eruption: a newly formed short loop that is pushed toward the surface, and an eruptive twisted long flux loop that carries the upward momentum. In the X2.2 flare, the former was manifested by the horizontal magnetic field enhancement on the surface \citep{2012ApJ...745L..17W}, and the latter was observed as a flux rope in corona that becomes part of the coronal mass ejection (CME) \citep{2011ApJ...738..167S}. There also exists some theoretical models that predict a sudden change in rotational motion of sunspots on the surface as a result of  coronal transients. In the simulation of \cite{2009ApJ...697.1529F}, the vorticity of the two spots at the feet of an emerging flux tube is enhanced in the same direction and with a similar magnitude for a short time, when coronal magnetic reconnection occurs during the flux emergence.

In this letter, we focus on the sudden photospheric motions associated with the 2011 February 15 X2.2 flare. We scrutinize the sudden change in rotational motion of the two main sunspots and the shear flow variation close to the flaring PIL. The spatiotemporal changes of the horizontal Lorentz force are analyzed quantitatively, and our main goal is to examine the possible relationship between the observed sudden motions and the flare-related Lorentz force changes.

\section{Observations and Data Processing}

Intensity images from the Helioseismic and Magnetic Imager \citep[HMI;][]{2012SoPh..275..229S} on board the Solar Dynamics Observatory were used to study the photospheric motions. The data with a pixel size of 0.5\arcsec\ and a cadence of 45 s are reconstructed from the profile of Fe~{\sc i} absorption line at 6173 \AA. The differential affine velocity estimator \citep[DAVE;][]{2006ApJ...646.1358S} was applied to the HMI intensity images to derive the photospheric flow field. The DAVE method used combines the advection equation and a differential feature tracking technique to detect flows. We used a window size of 19 pixels according to former studies \citep[e.g.,][]{2013SoPh..287..279L}, which is large enough to include structure information and small enough to have a good spatial resolution. The Fourier local correlation tracking method \citep[FLCT;][]{2008ASPC..383..373F} was applied separately to confirm the result. The FLCT spatial windowing parameter $\sigma$ was set to 7 pixels, corresponding to the same window size as in DAVE method. There are no optional parameters invoked in our use of FLCT. We also used the vector magnetograms supplied by the HMI team to investigate the vector field change. The vector data are derived from the observed Stokes parameters of the Fe~{\sc i} 6173~\AA\ line. The Stokes parameters are inverted with the Very Fast Inversion of the Stokes Vector \citep{2011SoPh..273..267B}. The 180$^{\circ}$ ambiguity is resolved using the minimum energy method \citep{1994SoPh..155..235M,2009SoPh..260...83L}.

The flare started at 01:44~UT, peaked at 01:56~UT, and ended at 02:06~UT in GOES 1--8~\AA\ flux. In order to study the sunspot rotation during the flare, regions of interests (ROIs) are defined with a threshold of 1300 G of the vertical magnetic field strength. In intensity images, these ROIs correspond to the umbral region of the sunspots. We show in Figure 1 the ROIs marked with p and f, which correspond to the proceding and following spots, respectively. We also select other two ROIs marked as s1 and s2, which are two  adjacent rectangular regions along the flaring PIL for the study of the shear flow. The change of horizontal Lorentz force $\delta \textbf{F}_h$ was computed using the 12 minute cadence vector magnetograms. \cite{2012SoPh..277...59F} formulated $\delta \textbf{F}_h$ at and below the photosphere as a surface integral reduced from a volume integral by using the Gauss' theorem (Eq. 18 of \citealt{2012SoPh..277...59F}):

\begin{equation}
\delta \textbf{F}_{h}=\frac{1}{4\pi}\int_{A_{ph}}dA\delta (B_{r}\textbf{B}_{h}) \ ,
\end{equation}

\noindent where $B_{r}$ is the photospheric vertical field and $\textbf{B}_{h}$ is the photospheric horizontal field. The torque ($ \textbf{T} = \textbf{r} \times \delta \textbf{F}_h $) of horizontal Lorentz force applied to the ROIs p and f can then be estimated, where the ``center-of-mass'' centroids of the ROIs are located based on the vertical magnetic field and are used as the rotation axis. We note that transient magnetic field changes could be induced by flare emissions \citep{1981ApJ...243L..99P,2003ApJ...599..615Q}; however, we found no transient change of the horizontal magnetic field in this event.

\section{Results}

The preceding (p) and following (f) spots have positive and negative polarity, respectively. The flow map of the p spot before the X2.2 flare at 01:27 UT is presented in Figure 2(a). The spot generally rotates in the clockwise direction at an average speed of 0.2 km~s$^{-1}$ in its boundary region. During the flare, the speed increases to around 0.8 km~s$^{-1}$ at 01:51 UT as shown in Figure 2(b). Thus the p spot rotates three times faster after the flare occurrence. The change of horizontal Lorentz force shown in Figure 2(c) indicates that the force acting on the volume at and below the photosphere during the flare is also in the clockwise direction, and may provide the driving force for the increased rate of spot rotation. Figures 2(d)--(f) display the maps of flows and Lorentz force in the f spot region, which show similar patterns of rotation and torque to those of the p spot. The centroid of the p(f) spot is marked with a green point in Figure 2(c(f)), and is used as the pivot to calculate the torque provided by the horizontal Lorentz force.

Based on the derived DAVE flow maps, we compute the flow vorticity at different times, and present the result in Figure 3. It can be seen that a negative vorticity dominates the umbral regions during the flare as shown in Figure 3(b), which is consistent with our flow tracking result that both the p and f spots undergo a sudden clockwise rotation during the flare.

We define the shear flow as the difference between the average velocity in the positive s2 and negative s1 regions in the direction parallel to the flaring PIL. The shear force is defined in a similar way based on the total Lorentz force in these regions. The shear force and the shear flow within the regions s1 and s2 (as denoted in Figure 1) are studied, and the result is shown in Figure 4(b).  The shear flow is about 0.2 km~s$^{-1}$ before 01:53 UT. At $\sim$01:55 UT the shear flow suddenly reverses its direction. Then it recovers to about 0.1 km~s$^{-1}$ at $\sim$02:07 UT and remains roughly constant afterward. The fact that there is a 50\% reduction of the shear flow after the flare is similar to the event studied by \cite{2009ApJ...690.1820T}. The sudden shear flow decrease is co-temporal with a $\sim 2.9 \times 10^{22}$ dyne of shear force, which is in the direction opposite to that of the initial shear flow. The shear force is close to zero at non-flaring periods. Therefore, the sudden decrease of shear flow speed is likely related to the horizontal Lorentz force.

Figure 4(c) shows the time profiles of rotational speed and torque of the preceding spot p. It rotates at about $6^{\circ}$~hr$^{-1}$ clockwise before the flare. At $\sim$01:49 UT, the rotational motion accelerates, and the angular speed reaches $37^{\circ}$~hr$^{-1}$ clockwise at $\sim$01:52 UT. The rotational speed then decreases to $\sim$1$^{\circ}$~hr$^{-1}$ clockwise in $\sim$5 minutes. This sudden motion results in about 3$^{\circ}$ clockwise rotation within 8 minutes. As a comparison, \cite{2012ApJ...744...50J} reported that several features on the outer edge of the penumbra of the p spot undergo a clockwise rotation from 20 hr before the flare to 1 hr after flare, with a speed of 1.8--5.1$^{\circ}$~hr$^{-1}$. Although we use a different method, our results are consistent with this previous study. The temporal evolution of the following spot f is displayed in Figure 4(d). The start time of the sudden clockwise rotation is $\sim$01:48 UT, and the angular speed reaches maximum at $\sim$01:51~UT co-temporal with the occurrence of the maximum torque. The overall property of the f region is similar to that of the p region. It is noticeable that the torque is only present during the flare, and hence it is likely responsible for the above sudden change in rotational motions during the flare.

Our observational results suggest that the horizontal Lorentz force is the driving force of the sudden photospheric motion. Specifically, the rotational speed of p and f spots increases $27^{\circ}$~hr$^{-1}$ in 225 s and 315 s, which results in an angular acceleration $\alpha$ of $5.8 \times 10^{-7}$~rad~s$^{-2}$ and $4.2 \times 10^{-7}$~rad~s$^{-2}$, respectively. In the s1 and s2 regions, the shear flow changes 0.4~km~s$^{-1}$ in 45~s, which corresponds to an acceleration of $1 \times 10^3$~cm~s$^{-2}$. As a more quantitative analysis, we compare these observed acceleration with those derived based on the measured torque $T$ due to the change of Lorentz force. For the spots p and f, we assume a geometry of rigidly rotating disk, for which $T = I \alpha$ and the moment of inertia $I = \frac{1}{2} \rho \pi h r^4$, where the density $\rho \approx 4 \times 10^{-7}$~g~cm$^{-3}$, depth $h \approx 250$~km (density scale height at the photosphere), and radius $r \approx 10$\arcsec. Since the measured $T \approx 3.9 \times 10^{30}$ ($2.5 \times 10^{30}$)~dyne~cm for the p (f) spot at 02:10 UT, the torque can produce an $\alpha$ of $8.8 \times 10^{-7}$ ($5.7 \times 10^{-7}$)~rad~s$^{-2}$, which is roughly comparable with the observation. Similarly, for the rectangular regions s1 and s2 (about 18\arcsec\ by 12\arcsec) with an estimated mass $m \approx 1.1 \times 10 ^{19}$~g, the measured shear force $\delta F_h =2.9 \times 10^{22}$~dyne  at 02:10 UT would cause an acceleration $\delta F_h / m \approx 2.6 \times 10^3$~cm~s$^{-2}$, which is also close to the observed acceleration. These results are summarized in Table 1. We caution that there is a large uncertainty in our calculation, which presumes a density scale height as the coherent depth of the rotational motion. Nevertheless, the acceleration due to the Lorentz force change is within the right order of magnitude to explain the observed changes in the rotation rate of spots and the shearing speed near the PIL.

\begin{table}[hb]
\caption{Comparison of Acceleration of Photospheric Regions} 
\centering 
\small
\begin{tabular}{c c c c} 
\hline\hline 
  &  $\alpha$ in p region & $\alpha$ in f region & acceleration in s1 \& s2 regions  \\ [0.5ex] 
\hline 
DAVE observed                & $5.8 \times 10^{-7}$~rad s$^{-2}$ & $4.2 \times 10^{-7}$~rad s$^{-2}$ & $1 \times10^{3} $~cm s$^{-2}$  \\
Estimated from $\delta F_h$ & $8.8 \times 10^{-7}$~rad s$^{-2}$ & $5.7 \times 10^{-7}$~rad s$^{-2}$ & $2.6 \times10^{3} $~cm s$^{-2}$  \\
\hline 
\end{tabular}
\end{table}

We further compare the initiation time of the sudden motions in different regions. The two spots p and f start the sudden rotation at almost the same time at $\sim$01:48 UT, while the sudden decrease of the shear flow occurs at $\sim$01:55~UT, i.e., 7 minutes later. Interestingly, the flare hard X-ray emission up to 100~keV has a first minor peak around 01:48~UT and has a main peak at 01:55~UT (see Figure~5(a)), the latter of which could be co-temporal with the peak of CME acceleration \citep{2008ApJ...673L..95T}. Therefore, this time gap of the above sudden motions might reflect the different response time of the different photospheric regions to the coronal field restructuring. Specifically, we surmise that the sudden change of the rotational motion of the sunspots at the feet of the flux rope could represent an immediate response to the flux rope eruption. In contrast, the delayed sudden shear flow change in the central flaring region could indicate a later effect of the magnetic field implosion process accompanying the rapid CME acceleration \citep[e.g.,][]{2011ApJ...727L..19L}.

\section{Summary and Discussion}

Using SDO/HMI intensity images with a high spatiotemporal resolution, we have carried out a detailed analysis of the evolution of photospheric flows associated with the 2011 February 15 X2.2 flare. We used the concept of vorticity to examine the flare-related sudden change in rotational motion of sunspots, which was considered difficult to measure. We have also investigated the sudden change of the shear flow around the flaring PIL. These results are discussed in the context of the change of the horizontal Lorentz force derived from SDO/HMI vector magnetograms. The main results are summarized as follows.

\begin{enumerate}
\item The two sunspots at the feet of the erupting flux rope show sudden acceleration of rotational motion during the flare. The fast clockwise rotation lasts $\sim$5 minutes and reaches a peak angular speed of $\sim$30$^{\circ}$~hr$^{-1}$.

\item  The region around the flaring PIL shows a sudden decrease of the shear flow velocity after the onset of the flare. The shear flow changes $\sim$0.4 km~s$^{-1}$ within 1 minute, and recovers to 0.1 km~s$^{-1}$ (half of the preflare value) in $\sim$10 minutes. The region is co-spatial with the horizontal magnetic field enhancement as studied in \cite{2012ApJ...745L..17W}.

\item  The horizontal Lorentz force may be the driving force of the sudden motions in the photosphere. The direction, magnitude, and spatiotemporal distribution of the force and the motion are consistent. The change of horizontal Lorentz force in the sheared regions is $\sim3 \times 10^{22}$ dyne, and the torque in each of the sunspot region is $\sim3 \times 10^{30}$~dyne cm. The change of Lorentz force provides the needed torque to drive such sudden motions.
\end{enumerate}

A major result is that we reveal, for the first time, that the rotational speed of sunspots suddenly changes about 4 minutes after the onset of the X2.2 flare, and that this motion could be driven by the change of the horizontal Lorentz force exerted on the photosphere. We note the simulation of \cite{2009ApJ...697.1529F}, in which the vorticity of two spots with opposite polarities and rotating in the same direction could be enhanced in that direction with a similar magnitude, when a coronal magnetic reconnection occurs during the emergence of the flux tube. In our study, the results of the X2.2 flare show a similar vorticity change in the regions of the two main sunspots with opposite polarities, which is co-temporal with the first minor hard X-ray peak soon after the flare initiation. We speculate that this rotational change of sunspots represents an immediate response to the flux rope eruption. Also notably, the sudden change of the shear flow in the central flare region occurs 7 minutes after the sunspot rotational change. The shear flow change is co-temporal with the main peak of the hard X-ray emission and CME acceleration. As momentum conservation implies that the upward impulse exerted on the erupting CME must be balanced by a downward impulse acting on the surface, it might be possible that the delayed shear flow change at the flare core region is a manifestation of the response of the low atmosphere to the magnetic field implosion. Obviously,  studies of more events are needed in order to provide further observational evidence for the impact of coronal field restructuring on the photosphere.

As a final note, \cite{2011ApJ...741L..35Z, 2013SoPh..284..315Z} discussed the properties of seismic sources of this event. It is interesting that the two sources are not located at the strong HXR sources near s1 and s2; instead, they are associated with two remote locations near the edge of p and f sunspots, where HXRs are hardly detected. The authors used the erupting flux rope model to explain the observations. We speculate that the sudden change in horizontal sunspot rotation may be associated with these two seismic sources.

\acknowledgments
We thank the SDO/HMI team for magnetic and intensity images, Dr.~P.~W.~Schuck for the DAVE code, Drs.~G.~H.~Fisher and B.~T.~Welsch for the FLCT code, and the referee for valuable comments that greatly improved the paper. This work was supported by NSF grants AGS 0839216 and AGS 0849453, and by NASA grants NNX13AF76G and NNX13AG13G.

\clearpage

\begin{figure}
\epsscale{1.0}
\plotone{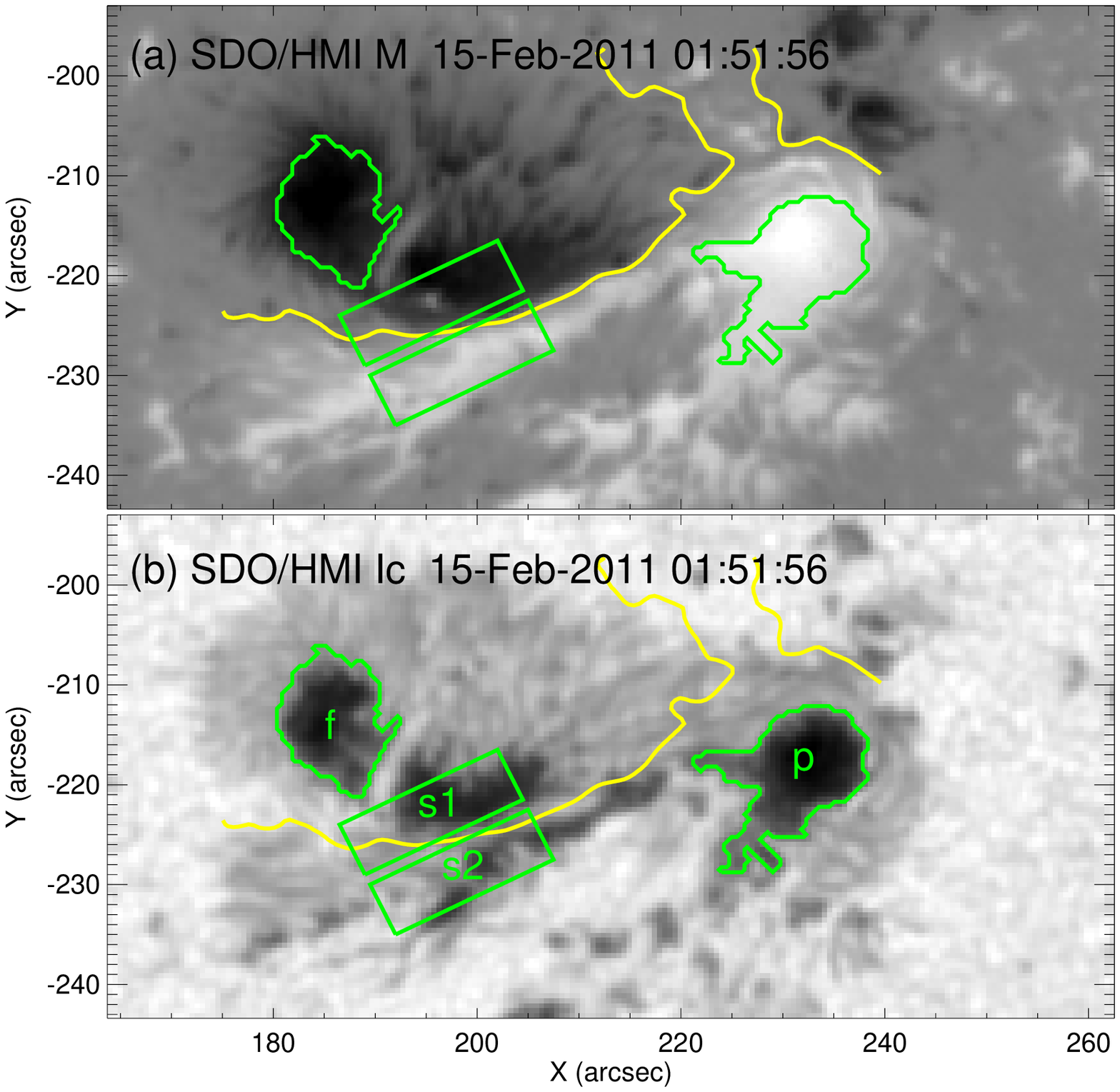}
\caption{Maps of AR 11158 with regions of interest marked by green contours. (a) SDO/HMI line-of-sight magnetogram. (b) SDO/HMI intensity image. The yellow curve represents the main flaring PIL.}
\end{figure}

\begin{figure}
\plotone{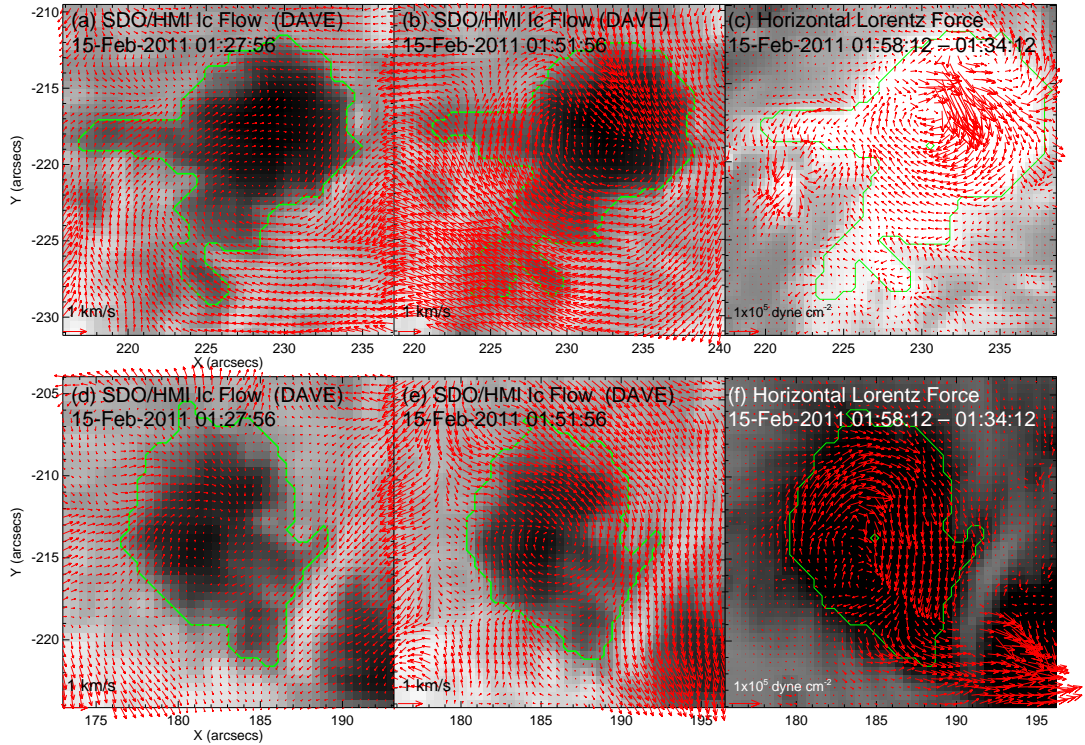}
\caption{DAVE flow maps of the region p before (a) and during (b) the flare. The change of horizontal Lorentz force during the flare is plotted in (c). DAVE flow maps of the region f before (d) and during (e) the flare. The change of horizontal Lorentz force during the flare is plotted in (f). The centroid of the p(f) spot is marked with a green point in (c(f)). To show the rotational motion better, the background constant translational motion is subtracted in the flow maps.}
\end{figure}

\begin{figure}
\begin{center}
\includegraphics[scale=.70]{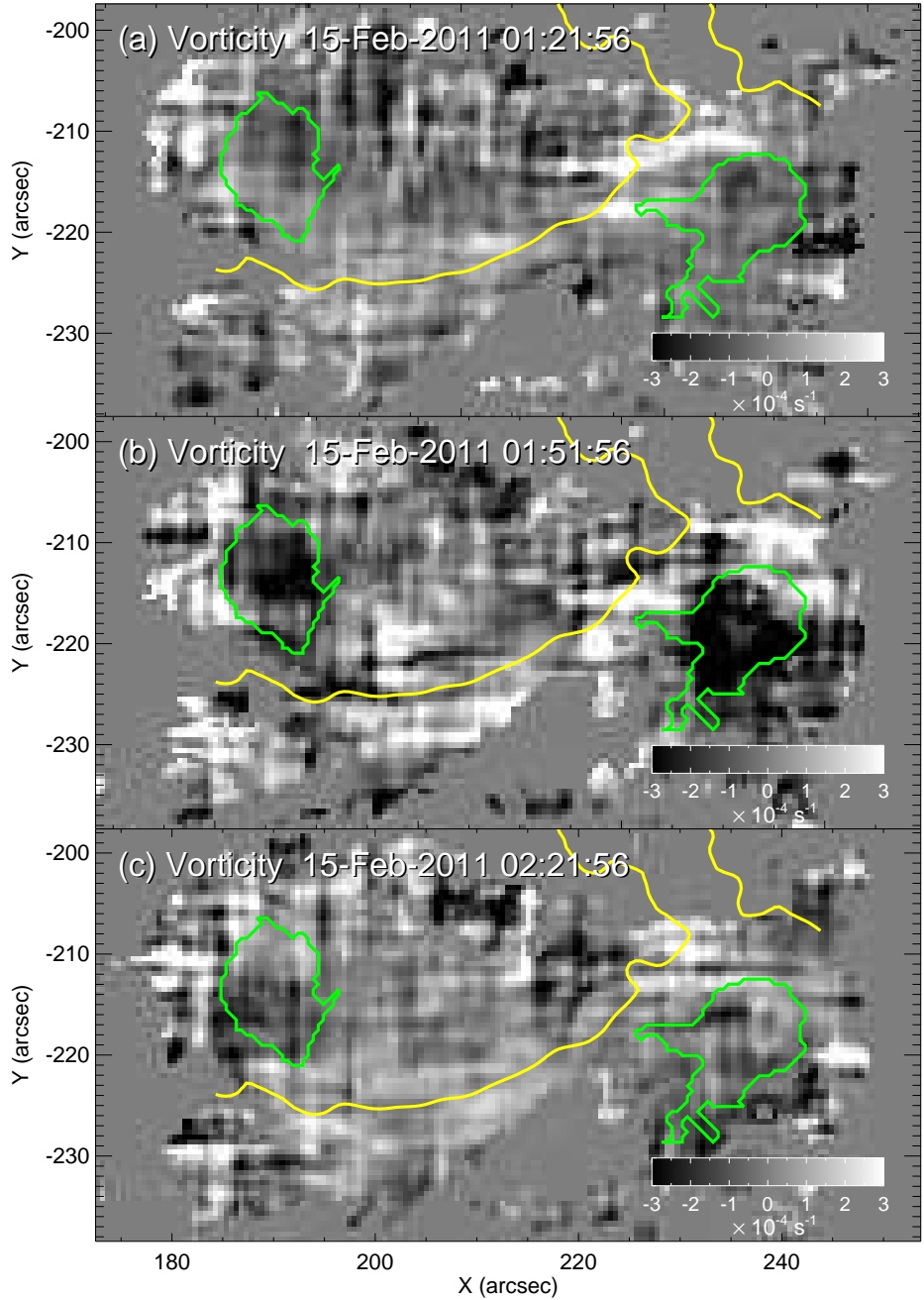}
\end{center}
\caption{Vorticity maps based on DAVE flows tracked based on SDO/HMI intensity images at different times. The sudden enhancement of negative vorticity in the sunspot areas (green contours) are co-spatial and co-tempoal with the flare.}
\end{figure}

\begin{figure}
\plotone{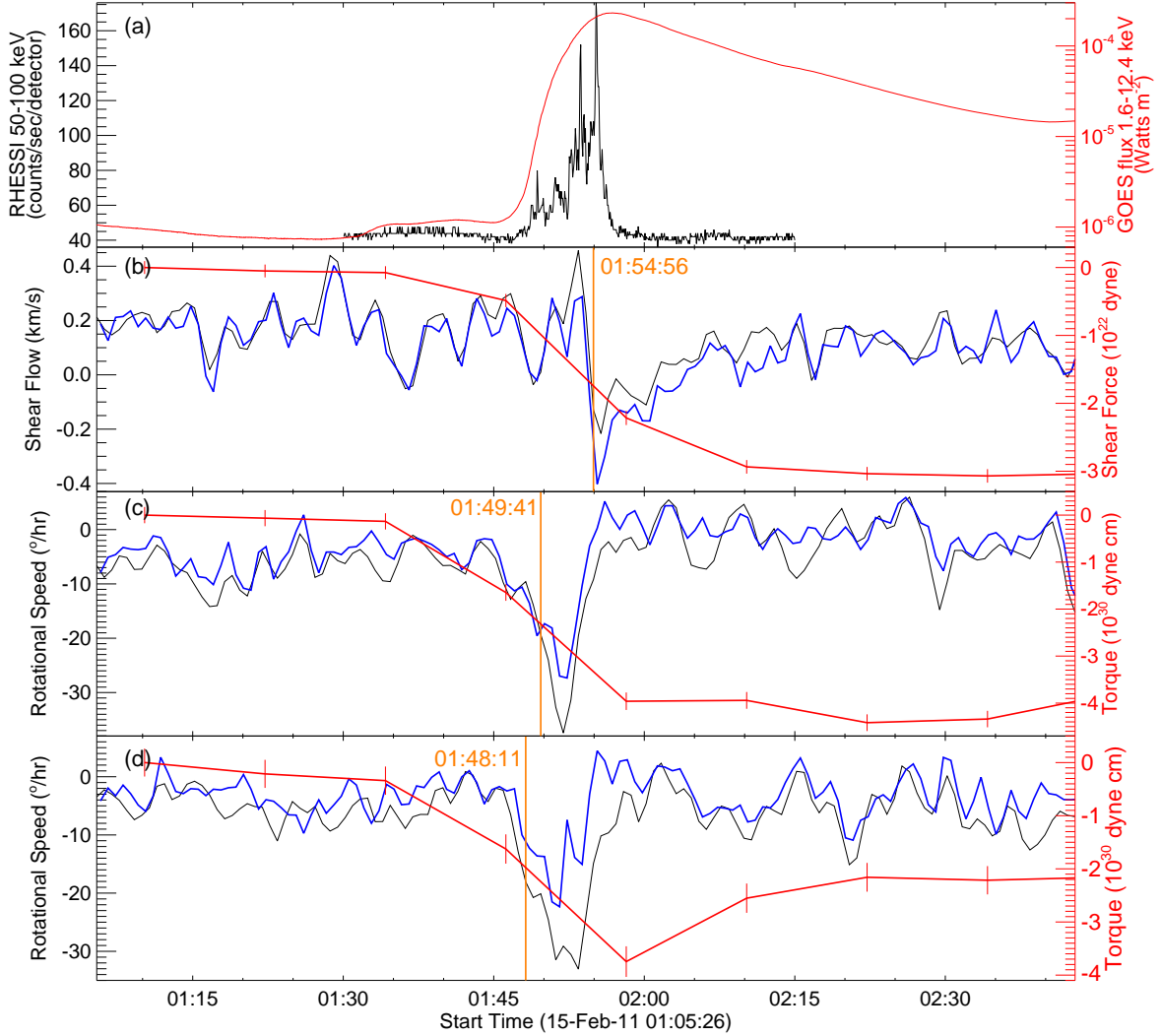}
\caption{Temporal evolution of the sudden motions. (a) The black curve represents RHESSI 50--100 keV HXR light curve. The red curve shows the GOES 1.6--12.4 keV flux. (b) Time profiles of the shear flow near PIL. The black and blue curves represent the mean velocity of the shear flow derived by DAVE and FLCT, respectively. The red curve shows the change of horizontal Lorentz force. (c) and (d) display the time profiles of the regions p and f, respectively. The black curves give the vorticity derived using DAVE flows, while the blue curves are from the FLCT result. The red curves show the torque provided by the change of horizontal Lorentz force. The orange vertical lines marked with time show the starting time of the sudden shear motion and rotations. The error bars of red curves indicate a 3$\sigma$ level.}
\end{figure}

\end{document}